\documentclass[%
 reprint,
 amsmath,amssymb,
 aps,
]{revtex4-2}

\usepackage{graphicx}
\usepackage{dcolumn}
\usepackage{bm}

\begin{document}

\preprint{APS/123-QED}

\title{Note on the Corrected Hawking Temperature of a Parametric Deformed Black Hole
with Influence of Quantum Gravity}

\author{Zhonghua Li}
\affiliation{Institute of Theoretical Physics, China West Normal University,Nanchong, Sichuan 637002, China}

\date{\today}

\begin{abstract}
\begin{description}
\item[Correspondence]
Correspondence should be addressed to Zhonghua Li; sclzh888@163.com.
\item[Data Availability]
All data used to support the findings of this study are included within the article. 
\item[Conflicts of Interest]
The authors declare that they have no conflicts of interest.
\end{description}
 In this paper, using Hamilton-Jacobi method, we discuss the tunnelling of fermions  when the dual influence of quantum gravity and the deformation of a parameterized black hole is taken into account. With the influence of the generalized uncertainty principle, there exists an offset around the standard Hawking temperature. We investigate a parametric deformed black hole and find that the corrected temperature is lower than the standard one,so there exists remnant of the black hole and the correction is not only determined by the mass of the emitted fermion,but also determined by the mass of the black hole and the deformation parameter.Under the dual influence of quantum gravity and deformation,the correction effect of quantum gravity is the  main influencing factor, while the correction effect of deformation parameter is secondary. The quantum gravitational correction factor is only determined by the mass of the emitted fermions, while the deformation correction factor is only determined by the mass of the black hole.

\end{abstract}

\maketitle


\section{\label{sec:level1}Introduction}
Hawking radiation near the event horizon of the black holes was found in the last century.
To analyze this phenomenon, researchers have made extensive studies. The usual method is by adopting the WKB approximation to calculate the imaginary part of the emitted particle's action and the tunnelling rate \cite{ETA1,ETA2}. The Hamilton-Jacobi method was first proposed in \cite{SP,SPS}. In this method, the action of the emitted particles satisfies the Hamilton-Jacobi equation. Taking into account the properties of the spacetime, one can carry out a separation of variables with the action $I=-\epsilon t+W(r)+\Phi(\theta,\phi)$.Then inserting the separated variables into the Hamilton-Jacobi equation and solving it, one can obtain the imaginary part. Extending this work to the tunnelling radiation of fermions, the standard Hawking temperatures of spherically symmetric and charged black holes were discussed in \cite{KM2}. Other work about fermions' tunnelling radiation is referred to \cite{KM3,LR,CJZ,QQJ,HCNVZ,BRM}.

 According to the theory of quantum gravity, there is  a minimal observable length \cite{PKTC,PKTC2,PKTC3,PKTC4}. This length can be used in the model of the generalized uncertainty principle (GUP)
\begin{eqnarray}
\Delta x \Delta p \geq \frac{\hbar}{2}\left[1+ \beta \Delta p^2\right],
\label{eq1.1}
\end{eqnarray}
\noindent where $\beta = \beta_0 (l^2_p /\hbar^2)$, $\beta_0 $ is a dimensionless parameter and $l_p$ is the Planck length. The derivation of the GUP is based on the modified fundamental commutation relations. Kempf et al. first modified commutation relations \cite{KMM} and got $\left[x_i,p_j\right]= i \hbar \delta_{ij}\left[1+ \beta p^2\right]$, where $x_i$ and $p_i$ are operators of position and momentum defined by
\begin{eqnarray}
x_i = x_{0i}, p_i = p_{0i} (1 + \beta p^2),
\label{eq1.2}
\end{eqnarray}

\noindent And here $x_{0i}$ and $p_{0i}$ satisfy the basic quantum mechanics commutation relations $\left[x_{0i},p_{0j}\right]= i \hbar \delta_{ij}$.

This generalization of the uncertainty principle based on the modified commutation relations plays an important role in quantum gravity. With consideration of the modifications, the cosmological constant problem was discussed and the finiteness of the constant was derived in \cite{CMOT}. Using a new form of GUP, the Unruh effect has been analyzed in \cite{BM06}. The quantum dynamics of the Friedmann-Robertson-Walker universe was got in \cite{BM1}. The related predictions on post-inflation preheating in cosmology were derived in \cite{CDAV}. Based on the modifications, the thermodynamics of the black holes were discussed again in \cite{KSY,XW,BJM,CL,LZ,CS,PLZY,JB} and the tunnelling radiation of scalar particles was studied in \cite{NS,LZH1}.

Alternative theories of gravity and the parameterized deviation approach allow black hole solutions to have
additional parameters beyond mass, charge, and angular momentum \cite{MLC}. Some extensions of general relativity have been proposed as alternative theories of gravity, with their corresponding black hole solutions. Recently, Johannsen and Psaltis introduced a parametric deviation approach \cite{JP}. This method can avoid some limitations of the original bumpy black hole approach \cite{CH,VH,VYS}. Among the parametric deformed solutions that emerged, Konoplya and Zhidenko proposed a Kerr-like solution, introducing a parametric deformation in the mass term,keeping the asymptotic behavior of the Kerr spacetime, but changing how the mass of the black hole influences the event horizon vicinity \cite{KZ}.

In this paper, we focus on the tunnelling radiation of fermions from a parametric deformed black hole, where the effects of quantum gravity are taken into account. That is ,we investigate the correction effect of Hawking temperature when the dual influence of quantum gravity and deformation of a parameterized black hole is taken into account. To incorporate the effects of quantum gravity, we first modify the Dirac equation in curved spacetime by the operators of position and momentum defined in \cite{KMM}. Then adopt the Hamilton-Jacobi method to get the imaginary parts of the action. By calculating, we want to know that the double effects of quantum gravity and deformation for a parameterized black hole and which one is the main influencing factor.

The rest is organized as follows. In the next section, to facilitate further discussion, we will review the generalized Dirac equation in curved spacetime. In Section 3, we investigate the tunnelling radiation of a  parametric deformed black hole. Section 4 is devoted to our conclusions. We use the natural units $G = c = \hbar = 1$, and signature $(-,+,+,+)$.
\section{Generalized Dirac equation in curved spacetime}

Based on the modified fundamental commutation relation  in \cite{KMM}, one can modify the Dirac equation in curved spacetime. According to Eq. (\ref{eq1.2}), the square of momentum operator is got as
\begin{eqnarray}
p^2 &=& p_i p^i  \nonumber \\
&=& -\hbar^2 \left[ {1 - \beta \hbar^2 \left( {\partial _j \partial ^j} \right)}
\right]\partial _i \cdot \left[ {1 - \beta \hbar^2 \left( {\partial ^j\partial _j } \right)}
\right]\partial ^i\nonumber \\
&\simeq & - \hbar ^2\left[ {\partial _i \partial ^i - 2\beta \hbar ^2
\left( {\partial ^j\partial _j } \right)\left( {\partial
^i\partial _i } \right)} \right].
\label{eq2.1}
\end{eqnarray}
\noindent Because of the small value of $\beta$, the higher order terms of $\beta $ can be neglected. According to the theory of quantum gravity, the generalized frequency is $\tilde \omega = E( 1 - \beta E^2)$,where $E$ is the energy operator and denoted as $ E = i \hbar \partial _t $. With consideration of the energy mass shell condition $ p^2 + m^2 = E^2 $, the generalized expression of the energy was got as \cite{NS}
\begin{eqnarray}
\tilde E = E[ 1 - \beta (p^2 + m^2)].
\label{eq2.3}
\end{eqnarray}
In curved spacetime, the Dirac equation is
\begin{eqnarray}
i\gamma^{\mu}\left(\partial_{\mu}+\Omega_{\mu}\right)\Psi+\frac{m}{\hbar}\Psi=0,
\label{eq2.4}
\end{eqnarray}
\noindent where $\Omega _\mu \equiv\frac{i}{2}\omega _\mu\, ^{a b} \Sigma_{ab}$, $\omega _\mu\, ^{ab}$ is the spin connection defined  by the tetrad $e^\lambda\,_b$ and the ordinary connection
\begin{eqnarray}
\omega_\mu\,^a\,_b=e_\nu\,^a e^\lambda\,_b \Gamma^\nu_{\mu\lambda} -e^\lambda\,_b\partial_\mu e_\lambda\,^a.
\label{eq2.5}
\end{eqnarray}
\noindent The Latin indices live in the flat metric $\eta_{ab}$ while Greek indices are raised and lowered by the curved metric $g_{\mu\nu}$. The tetrad can be constructed from
\begin{eqnarray}
g_{\mu\nu}= e_\mu\,^a e_\nu\,^b \eta_{ab},\nonumber \\
\eta_{ab}= g_{\mu\nu} e^\mu\,_a e^\nu\,_b, \nonumber \\
e^\mu\,_a e_\nu\,^a=\delta^\mu_\nu,\nonumber \\
e^\mu\,_a e_\mu\,^b = \delta_a^b.
\label{eq2.6}
\end{eqnarray}
\noindent In equation (\ref{eq2.4}), $\Sigma_{ab}$ is the Lorentz spinor generator defined by
\begin{eqnarray}
\Sigma_{ab}= \frac{i}{4}\left[ {\gamma ^a ,\gamma^b} \right], \nonumber \\
\{\gamma ^a ,\gamma^b\}= 2\eta^{ab}.
\label{eq2.7}
\end{eqnarray}
\noindent Then one can construct $\gamma^\mu$'s in the curved spacetime as
\begin{eqnarray}
\gamma^\mu = e^\mu\,_a \gamma^a, \nonumber \\
\left\{ {\gamma ^\mu,\gamma ^\nu } \right\} = 2g^{\mu \nu }.
\label{eq2.8}
\end{eqnarray}
\noindent To get the generalized Dirac equation in the curved spacetime, one can rewrite Eq. (\ref{eq2.4}) as
\begin{eqnarray}
-i\gamma^{0}\partial_{0}\Psi=\left(i\gamma^{i}\partial_{i}+i\gamma^{\mu}\Omega_{\mu}+\frac{m}{\hbar}\right)\Psi.
\label{eq2.9}
\end{eqnarray}
\noindent Using Eqs. (\ref{eq2.1}), (\ref{eq2.3}) and (\ref{eq2.9}) and neglecting the higher order terms of $\beta$, one can  get
\begin{eqnarray}
-i\gamma^{0}\partial_{0}\Psi=\left(i\gamma^{i}\partial_{i}+i\gamma^{\mu}\Omega_{\mu}+\frac{m}{\hbar}\right)\nonumber \\
\cdot \left(1+\beta\hbar^{2}\partial_{j}\partial^{j}-\beta m^{2}\right)\Psi,
\label{eq2.10}
\end{eqnarray}
\noindent which is rewritten as
\begin{eqnarray}
\left[i\gamma^{0}\partial_{0}+i\gamma^{i}\partial_{i}\left(1-\beta m^{2}\right)
+i\gamma^{i}\beta\hbar^{2}\left(\partial_{j}\partial^{j}\right)\partial_{i} \right.\nonumber \\ +\frac{m}{\hbar}\left(1+\beta\hbar^{2}\partial_{j}\partial^{j}-\beta m^{2}\right)\nonumber \\
\left.+i\gamma^{\mu}\Omega_{\mu}\left(1+\beta\hbar^{2}\partial_{j}\partial^{j}-\beta m^{2}\right)\right]\Psi= 0.
\label{eq2.11}
\end{eqnarray}
\noindent Thus, the generalized Dirac equation is derived. In the following sections, we adopt Eq. (\ref{eq2.11}) to describe fermion tunnelling of a parametric deformed black hole.

\section{Fermions' tunnelling of a parametric deformed black hole}

In this section, we investigate fermions' tunnelling from the event horizon of a parametric deformed black hole.
 To do so, we choose a static
spherically symmetric black hole metric [7]
\begin{eqnarray}
  ds^2=-F(r)dt^2+\frac{1}{G(r)}dr^2 \nonumber \\
  +r^2d\theta^2+r^2sin^2\theta d\phi^2,
\label{eq4.1}
\end{eqnarray}
where
\begin{eqnarray}
F(r)=G(r)=1 -\frac{2M}{r} - \frac{\eta}{r^3},
\label{eq4.11a}
\end{eqnarray}
$0\leq\phi\leq2\pi,0\leq\theta\leq\pi,0\leq r < \infty, \eta $ is the deformation parameter of the black hole, and $M$ is the mass of the black hole. For the sake of simplicity in the discussion, we take $ 0\leq\eta \leq 1 $ and choose the real root of the event horizon located at \cite{MLC}

\begin{eqnarray}
  r_{+}=\frac{2M}{3}+(\frac{27\eta+16M^3}{54}+A(\eta))^{\frac{1}{3}}\nonumber\\
  +(\frac{27\eta+16M^3}{54}-A(\eta))^{\frac{1}{3}},
\label{eq4.a1}
\end{eqnarray}
where
\begin{eqnarray}
A(\eta)=((\frac{27\eta+16M^3}{54})^2-\frac{64M^6}{729})^{\frac{1}{2}},
\label{eq4.a11}
\end{eqnarray}
\noindent If the deformation parameter $\eta = 0$,the event horizon $r_{+}=2M$, it reduces to the event horizon of Schwarzchild black hole. Neglecting the higher order terms of $\eta$, we get
\begin{eqnarray}
 r_{+}=2M+\frac{3}{4M^2}\eta.
\label{eq4.a111}
\end{eqnarray}
\noindent Here we only investigate the spin-up state. Assume that the wave function of the fermion in spin-up state is

\begin{eqnarray}
\Psi=\left(\begin{array}{c}
A\\
0\\
B\\
0
\end{array}\right)\exp\left(\frac{i}{\hbar}I\left(t,r,\theta , \phi \right)\right).
\label{eq4.6}
\end{eqnarray}

\noindent The tetrad is easily constructed as

\begin{eqnarray}
e_\mu\,^a = \rm{diag}\left(\sqrt F, 1/\sqrt G, \sqrt {g^{\theta\theta}},\sqrt {g^{\phi\phi}} \right).
\label{eq4.7}
\end{eqnarray}

\noindent The gamma matrices take on the form as

\begin{eqnarray}
\gamma^{t}=\frac{1}{\sqrt{F\left(r\right)}}\left(\begin{array}{cc}
i & 0\\
0 & -i
\end{array}\right), &  & \gamma^{\theta}=\sqrt{g^{\theta\theta}}\left(\begin{array}{cc}
0 & \sigma^{1}\\
\sigma^{1} & 0
\end{array}\right),\nonumber \\
\gamma^{r}=\sqrt{G\left(r\right)}\left(\begin{array}{cc}
0 & \sigma^{3}\\
\sigma^{3} & 0
\end{array}\right), &  & \gamma^{\phi}=\sqrt{g^{\phi\phi}}\left(\begin{array}{cc}
0 & \sigma^{2}\\
\sigma^{2} & 0
\end{array}\right).
\label{eq4.8}
\end{eqnarray}

\noindent In the above equations, $g^{\theta\theta} = 1 / r^2$,  $g^{\phi\phi} = 1 / r^2 sin^2\theta $. Inserting the wave function and the gamma matrices into the generalized Dirac equation , we get

\begin{eqnarray}
-iA\frac{1}{\sqrt{F}}\partial_{t}I-B\left(1-\beta m^{2}\right)\sqrt{G}\partial_{r}I \nonumber\\
-Am\beta\left[g^{rr}\left(\partial_{r}I\right)^{2}+
g^{\theta\theta}\left(\partial_{\theta}I\right)^{2}+g^{\phi\phi}\left(\partial_{\phi}I\right)^{2}\right]\nonumber\\
+B\beta\sqrt{G}\partial_{r}I\left[g^{rr}\left(\partial_{r}I\right)^{2}+ g^{\theta\theta}\left(\partial_{\theta}I\right)^{2} + g^{\phi\phi}\left(\partial_{\phi}I\right)^{2}\right] \nonumber\\
+Am\left(1-\beta m^{2}\right) = 0,\nonumber\\
\label{eq4.9}
\end{eqnarray}

\begin{eqnarray}
iB\frac{1}{\sqrt{F}}\partial_{t}I-A\left(1-\beta m^{2}\right)\sqrt{G}\partial_{r}I \nonumber\\
-Bm\beta\left[g^{rr}\left(\partial_{r}I\right)^{2}+
g^{\theta\theta}\left(\partial_{\theta}I\right)^{2}+g^{\phi\phi}\left(\partial_{\phi}I\right)^{2}\right]\nonumber\\
+A\beta\sqrt{G}\partial_{r}I\left[g^{rr}\left(\partial_{r}I\right)^{2}+g^{\theta\theta}\left(\partial_{\theta}I\right)^{2} + g^{\phi\phi}\left(\partial_{\phi}I\right)^{2}\right] \nonumber\\
+Bm\left(1-\beta m^{2}\right) = 0,\nonumber\\
\label{eq4.10}
\end{eqnarray}

\begin{eqnarray}
A\left\{-\left(1-\beta m^{2}\right)\sqrt{g^{\theta\theta}} \partial_{\theta}I \right.\nonumber\\
\left.+\beta\sqrt{g^{\theta\theta}}\partial _{\theta}I\left[g^{rr}\left(\partial_{r}I\right)^{2}+
g^{\theta\theta}\left(\partial_{\theta}I\right)^{2}+g^{\phi\phi}\left(\partial_{\phi}I\right)^{2}\right]\right.\nonumber\\
-i\left(1-\beta m^{2}\right)\sqrt{g^{\phi\phi}}\partial_{\phi}I \nonumber\\
\left.+i\beta\sqrt{g^{\phi\phi}}\partial _{\phi}I\left[g^{rr}\left(\partial_{r}I\right)^{2}+
g^{\theta\theta}\left(\partial_{\theta}I\right)^{2}+g^{\phi\phi}\left(\partial_{\phi}I\right)^{2}\right]\right\} = 0,\nonumber\\
\label{eq4.11}
\end{eqnarray}

\begin{eqnarray}
B\left\{-\left(1-\beta m^{2}\right)\sqrt{g^{\theta\theta}} \partial_{\theta}I \right.\nonumber\\
\left.+\beta\sqrt{g^{\theta\theta}}\partial _{\theta}I\left[g^{rr}\left(\partial_{r}I\right)^{2}+
g^{\theta\theta}\left(\partial_{\theta}I\right)^{2}+g^{\phi\phi}\left(\partial_{\phi}I\right)^{2}\right]\right.\nonumber\\
-i\left(1-\beta m^{2}\right)\sqrt{g^{\phi\phi}}\partial_{\phi}I \nonumber\\
\left.+i\beta\sqrt{g^{\phi\phi}}\partial _{\phi}I\left[g^{rr}\left(\partial_{r}I\right)^{2}+
g^{\theta\theta}\left(\partial_{\theta}I\right)^{2}+g^{\phi\phi}\left(\partial_{\phi}I\right)^{2}\right]\right\} = 0. \nonumber\\
\label{eq4.12}
\end{eqnarray}

\noindent It is difficult to solve the action $I$ from the above equations.  Considering the properties of the metric (\ref{eq4.1}), we carry out the separation of variables as

\begin{eqnarray}
I=-\epsilon t+ W\left(r \right) + \Phi(\theta,\phi),
\label{eq4.14}
\end{eqnarray}

 \noindent where $\epsilon$  is the energy of the radiation particle. We first observe Eqs. (\ref{eq4.11}) and (\ref{eq4.12}) and find that they are irrelevant to $A$ and $B$ and can be reduced to the same equation. Inserting Eq. (\ref{eq4.14}) into Eqs. (\ref{eq4.11}) and (\ref{eq4.12}) yields

\begin{eqnarray}
\left(\sqrt{ g^{\theta\theta}}\partial _ {\theta} \Phi + i\sqrt {g^{\phi\phi}}\partial _ {\phi}\Phi\right) \nonumber\\
\left[1 - \beta m^2 -\beta g^{rr}
(\partial _r W)^2 - \beta g^{\theta\theta}(\partial _ {\theta}\Phi)^2 - \beta  g^{\phi\phi}(\partial _ {\phi}\Phi)^2\right] = 0. \nonumber\\
\label{eq3.13}
\end{eqnarray}

\noindent In the above equation, the summation of factors in the square brackets can not be zero. Therefore, it should be

\begin{eqnarray}
\sqrt{ g^{\theta\theta}}\partial _ {\theta}\Phi + i\sqrt {g^{\phi\phi}}\partial _ {\phi}\Phi = 0,
\label{eq3.14}
\end{eqnarray}

\noindent This implies

\begin{eqnarray}
g^{\theta\theta}\left(\partial_{\theta}\Phi\right)^{2}+ g^{\phi\phi}\left(\partial_{\phi}\Phi\right)^{2} =0.
\label{eq4.13}
\end{eqnarray}
which yields a complex function solution (other than the trivial constant solution) of $\Phi$. However, this solution has no contribution to the tunnelling rate. Therefore, we will not consider its contribution in the calculation.
\noindent Now our interest is the first two equations which determine the Hawking temperature of the black hole.
\noindent  Inserting Eq. (\ref{eq4.14}) into Eqs. (\ref{eq4.9}) and (\ref{eq4.10}) and canceling $A$ and $B$ yields

\begin{eqnarray}
A_6\left( {\partial _r W} \right)^6 + A_4\left({\partial _r W} \right)^4 +A_2\left( {\partial _r W} \right)^2 + A_0 = 0,
\label{eq4.15}
\end{eqnarray}

\noindent where
\begin{eqnarray}
A_6 & = & \beta^{2}G^{3}F,\nonumber \\
A_4 & = & \beta G^{2}F\left(m^{2}\beta -2\right),\nonumber \\
A_2 & = & GF\left[\left(1-\beta m^{2}\right)^{2}+2\beta m^{2}\left(1-m^{2}\beta\right)\right],\nonumber \\
A_0 & = & -m^{2}\left(1-\beta m^{2}\right)^{2}F- \epsilon^{2}.
\label{eq4.16}
\end{eqnarray}

\noindent Neglecting the higher order terms of $\beta$ and $\eta$ and considering Einstein mass-energy relation $\epsilon=m $,solving Eq. (\ref{eq4.15}) at the event horizon, we get the solution of $W$. Thus, the imaginary part of $W$ is

\begin{eqnarray}
Im W_{\pm} & = &\pm\int dr\sqrt{\frac{m^{2}F+\epsilon^{2}}{GF}} \left(1+\beta m^{2}+\beta\frac{\epsilon^{2}}{F}\right)\nonumber \\
 & = & \pm \pi B_0\left( 1+\beta \cdot B_1\left( 1+\eta \cdot B_2\right)\right),
\label{eq4.17}
\end{eqnarray}
\noindent where
\begin{eqnarray}
B_0 & = & \epsilon r_{+}^{4}(1+m^{2})(3r_{+}-4M)= m r_{+}^{4}(1+m^{2})(3r_{+}-4M),\nonumber \\
B_1 & = & \frac{m^2 + 12 \epsilon^2}{2(1+m^2)}=\frac{13m^2}{2(1+m^2)},\nonumber \\
B_2 & = & \frac{7m^2-72 \epsilon ^2}{8M^3(m^2+12 \epsilon ^2)}= - \frac{65}{104M^3},
\end{eqnarray}

In the above equation, the +/- sign corresponds to the outgoing / ingoing wave,$F=G=1 -\frac{2M}{r} - \frac{\eta}{r^3}$.
 Thus, the tunnelling rate of the fermion crossing the horizon is
\begin{eqnarray}
\Gamma & = & \frac{P(emission)}{P(absorption)}=\frac{exp(-2Im I_{+})}{exp(-2Im I_{-})} \nonumber \\
 & = & \frac{exp(-2Im W_{+}-2Im \Phi)}{exp(-2Im W_{-}-2Im \Phi)} \nonumber \\
 & = & exp[ - 4 \pi B_0\left( 1+\beta \cdot B_1\left( 1+\eta \cdot B_2\right)\right)]
\label{eq4.18}
\end{eqnarray}
Then the Boltzmann factor with Hawking temperature:
\begin{eqnarray}
T & = & \frac{1}{4 \pi B_0\left( 1+\beta \cdot B_1\left( 1+\eta \cdot B_2\right)\right)} \nonumber\\
  & = & T_{0} (1-\beta \cdot B_1\left( 1+\eta \cdot B_2\right)),
\label{eq4.19}
\end{eqnarray}
where $T_{0}=1/4 \pi B_0 $ is the standard Hawking temperature of the black hole.It is shown that the corrected temperature is lower than the standard one. The correction is not only determined by the mass of the emitted fermion $m$,but also determined by the mass of the black hole $M$ and the deformation parameter $\eta$. Clearly,we can also find easily that under the dual influence of quantum effects and deformation,the  correction effect of quantum gravity is the main influencing factor, while the correction effect of deformation parameter is secondary. Moreover, we find that the quantum gravity correction factor $B_1$ is only determined by the mass of the emitted fermion $m$,while the deformation correction factor $B_2$ is only determined by the mass of the black hole $M$. So with consideration of the dual influence of quantum gravity and deformation , the corrected Hawking temperature always exists and is lower than the standard one. that is , with the the evaporation proceeds, the Hawking temperature decreases, and the black hole finally reaches an equilibrium state, there exists remnant of the black hole.

\section{Conclusions}

In this paper, using Hamilton-Jacobi method, we discussed the tunnelling of fermions  when the dual influence of quantum gravity and the deformation of a parameterized black hole is taken into account. Taking into account the influence of quantum gravity, one can modified the Dirac equation in curved spacetime by the modified fundamental commutation relations. Then the tunnelling radiation of fermions from the event horizon of a parametric deformed black hole was investigated. The corrected Hawking temperatures were got. We found that

(i) the corrected temperature is lower than the standard one and that is , with the the evaporation proceeds, the Hawking temperature decreases, and the black hole finally reaches an equilibrium state, there exists remnant of the black hole;

(ii) the correction is not only determined by the mass of the emitted fermion $m$,but also determined by the mass of the black hole $M$ and the deformation parameter $\eta$;

(iii) under the dual influence of quantum effects and deformation,the  correction effect of quantum gravity is the main influencing factor, while the correction effect of deformation parameter is secondary;

(iv) the quantum gravity correction factor $B_1$ is only determined by the mass of the emitted fermion $m$,while the deformation correction factor $B_2$ is only determined by the mass of the black hole $M$.

\begin{acknowledgments}
This work is supported by the Fundamental Research Funds of China West Normal University (13C009).
\end{acknowledgments}



\begin{thebibliography}{0}%
\makeatletter
\providecommand \@ifxundefined [1]{%
 \@ifx{#1\undefined}
}%
\providecommand \@ifnum [1]{%
 \ifnum #1\expandafter \@firstoftwo
 \else \expandafter \@secondoftwo
 \fi
}%
\providecommand \@ifx [1]{%
 \ifx #1\expandafter \@firstoftwo
 \else \expandafter \@secondoftwo
 \fi
}%
\providecommand \natexlab [1]{#1}%
\providecommand \enquote  [1]{``#1''}%
\providecommand \bibnamefont  [1]{#1}%
\providecommand \bibfnamefont [1]{#1}%
\providecommand \citenamefont [1]{#1}%
\providecommand \href@noop [0]{\@secondoftwo}%
\providecommand \href [0]{\begingroup \@sanitize@url \@href}%
\providecommand \@href[1]{\@@startlink{#1}\@@href}%
\providecommand \@@href[1]{\endgroup#1\@@endlink}%
\providecommand \@sanitize@url [0]{\catcode `\\12\catcode `\$12\catcode
  `\&12\catcode `\#12\catcode `\^12\catcode `\_12\catcode `\%12\relax}%
\providecommand \@@startlink[1]{}%
\providecommand \@@endlink[0]{}%
\providecommand \url  [0]{\begingroup\@sanitize@url \@url }%
\providecommand \@url [1]{\endgroup\@href {#1}{\urlprefix }}%
\providecommand \urlprefix  [0]{URL }%
\providecommand \Eprint [0]{\href }%
\providecommand \doibase [0]{https://doi.org/}%
\providecommand \selectlanguage [0]{\@gobble}%
\providecommand \bibinfo  [0]{\@secondoftwo}%
\providecommand \bibfield  [0]{\@secondoftwo}%
\providecommand \translation [1]{[#1]}%
\providecommand \BibitemOpen [0]{}%
\providecommand \bibitemStop [0]{}%
\providecommand \bibitemNoStop [0]{.\EOS\space}%
\providecommand \EOS [0]{\spacefactor3000\relax}%
\providecommand \BibitemShut  [1]{\csname bibitem#1\endcsname}%
\let\auto@bib@innerbib\@empty
\end{thebibliography}%


\nocite{*}
\begin{thebibliography}{0}

\bibitem{ETA1}
 E. T. Akhmedov, V. Akhmedova, T. Pilling, and D. Singleton,
``Thermal radiation of various gravitational backgrounds,'' \emph{International Journal of Modern Physics A}, vol. 22, no. 8-9, pp. 1705-1715, 2007.

\bibitem{ETA2}
B. D. Chowdhury, ``Problems with tunneling of thin shells from
black holes,'' \emph{Pramana}, vol. 70, no. 1, pp. 3-26, 2008.

\bibitem{SP}
K. Srinivasan and T. Padmanabhan, ``Particle production and
complex path analysis,'' \emph{Physical Review D}, vol. 60, no. 2, Article
ID 024007, 20 pages, 1999.

\bibitem{SPS}
S. Shankaranarayanan, T. Padmanabhan, and K. Srinivasan,
``Hawking radiation in different coordinate settings: complex
paths approach,'' \emph{Classical and Quantum Gravity}, vol. 19, no. 10,
pp. 2671-2687, 2002.

\bibitem{KM2}
R. Kerner and R. B. Mann, ``Fermions tunnelling from black
holes,'' \emph{Classical and Quantum Gravity}, vol. 25, no. 9, Article ID
095014, 2008.

\bibitem{KM3}
R. Kerner and R. B. Mann, ``Charged fermions tunnelling from
Kerr-Newman black holes,'' \emph{Physics Letters B}, vol. 665, no. 4, pp.
277-283, 2008.

\bibitem{LR}
R. Li and J.-R. Ren, ``Dirac particles tunneling from BTZ black
hole,''\emph{Physics Letters B}, vol. 661, no. 5, pp. 370-372, 2008.

\bibitem{CJZ}
 D.-Y. Chen, Q.-Q. Jiang, and X.-T. Zu, ``Hawking radiation of
Dirac particles via tunnelling from rotating black holes in de
Sitter spaces,'' \emph{Physics Letters B}, vol. 665, no. 2-3, pp. 106-110,
2008.

\bibitem{QQJ}
Q.-Q. Jiang, ``Dirac particle tunneling from black rings,'' \emph{Physical Review D}, vol. 78, no. 4, Article ID 044009, 8 pages, 2008.

\bibitem{HCNVZ}
 R. D. Criscienzo and L. Vanzo, ``Fermion tunneling from
dynamical horizons,'' \emph{Europhysics Letters}, vol. 82, no. 6, Article
ID 60001, 2008.

\bibitem{BRM}
B. R. Majhi, `` Fermion tunneling beyond semiclassical approximation,'' \emph{Physical Review D}, vol. 79, Article ID 044005, 10 pages, 2009.

\bibitem{PKTC}
P. K. Townsend, ``Small-scale structure of spacetime as the
origin of the gravitational constant,'' \emph{Physical Review D}, vol. 15,
no. 10, pp. 2795-2801, 1977.

\bibitem{PKTC2}
K. Konishi, G. Paffuti, and P. Provero, ``Minimum physical
length and the generalized uncertainty principle in string
theory,'' \emph{Physics Letters B}, vol. 234, no. 3, pp. 276-284, 1990.

\bibitem{PKTC3}
 L. J. Garay, ``Quantum gravity and minimum length,'' \emph{International Journal of Modern Physics A}, vol. 10, no. 2, pp. 145-165,
1995.

\bibitem{PKTC4}
G. Amelino-Camelia, ``Relativity in spacetimes with short-distance structure governed by an observer-independent
(Planckian) length scale,'' \emph{International Journal of Modern
Physics D}, vol. 11, no. 1, pp. 35-59, 2002.

\bibitem{KMM}
A. Kempf, G. Mangano, and R. B. Mann, ``Hilbert space
representation of the minimal length uncertainty relation,''
\emph{Physical Review D}, vol. 52, no. 2, pp. 1108-1118, 1995.

\bibitem{CMOT}
L. N. Chang, D. Minic, N. Okamura, and T. Takeuchi, ``Effect of
the minimal length uncertainty relation on the density of states
and the cosmological constant problem,'' \emph{Physical Review D}, vol.
65, no. 12, Article ID 125028, 7 pages, 2002.

\bibitem{BM06}
B. R. Majhi and E. C. Vagenas, ``Modified dispersion relation,
photon’s velocity, and Unruh effect,'' \emph{Physics Letters B}, vol. 725,
no. 4-5, pp. 477-480, 2013.

\bibitem{BM1}
M. V. Battisti and G. Montani, ``The big-bang singularity in
the framework of a generalized uncertainty principle,'' \emph{Physics
Letters B}, vol. 656, no. 1-3, pp. 96-101, 2007.

\bibitem{CDAV}
W. Chemissany, S. Das, A. F. Ali, and E. C. Vagenas, ``Effect of the
generalized uncertainty principle on post-inflation preheating,''
\emph{Journal of Cosmology and Astroparticle Physics}, vol. 2011, no. 12,
article 17, 2011.

\bibitem{KSY}
W. Kim, E. J. Son, and M. Yoon, ``Thermodynamics of a black
hole based on a generalized uncertainty principle,'' \emph{Journal of
High Energy Physics}, vol. 2008, no. 1, article 35, 2008.

\bibitem{XW}
L. Xiang and X. Q. Wen, ``A heuristic analysis of black hole thermodynamics with generalized uncertainty principle,'' \emph{Journal of High Energy Physics}, vol. 2009, no. 10, article 46, 2009.

\bibitem{BJM}
A. Bina, S. Jalalzadeh, and A. Moslehi, ``Quantum black hole
in the generalized uncertainty principle framework,'' \emph{Physical
Review D}, vol. 81, no. 2, Article ID 023528, 7 pages, 2010.

\bibitem{CL}
D.-Y. Chen and Z.-H Li, ``Remarks on remnants by fermions’ tunnelling from
black strings,'' \emph{Advances in High Energy Physics},Volume 2014, Article ID 620157, 9 pages,2014.

\bibitem{LZ}
Z.-H Li and L.-M Zhang, ``Fermions tunnelling from black string and Kerr AdS
black hole with consideration of quantum gravity,'' \emph{International Journal of Theoretical Physics},Volume 54, no 6, pp. 1739-2118 ,2015.

\bibitem{CS}
S. Chakraborty and S. Saha, ``Quantum tunnelling for Hawking radiation from both static and dynamic black holes,'' \emph{Advances in High Energy Physics},Volume 2014, Article ID 168487,2014.

\bibitem{PLZY}
 J.Pu, K. Lin, X.-T Zu, S.-Z Yang, ``Modified Fermions Tunneling Radiation from Non-stationary, Axially Symmetric Kerr Black Hole,'' \emph{Advances in High Energy Physics},Volume 2019, Article ID 5864042,2019.

\bibitem{JB}
W. Javed and R. Babar, ``Fermions Tunneling and Quantum Corrections for Quintessential Kerr-Newman-AdS Black Hole,'' \emph{Advances in High Energy Physics},Volume 2019, Article ID 2759641,2019.

\bibitem{NS}
K. Nozari and S. Saghafi,``Natural cutoffs and quantum tunneling from black hole horizon,''\emph{Journal of High Energy Physics},2012, Article number 5,2012.

\bibitem{LZH1}
Z.-H Li,``Scalar Particles Tunneling Radiation in the Demianski-Newman
Spacetime with Influences of Quantum Gravity,'' \emph{Advances in High Energy Physics},Volume 2020, Article ID 7549728, 5 pages,2020.

\bibitem{MLC}
R. B. Magalhaes, L. C. S. Leite, L. C. B. Crispino,``Absorption by deformed black holes,''\emph{Physics Letters B} , vol.805,
Article ID 135418, 2020.

\bibitem{JP}
T. Johannsen and D. Psaltis,``Metric for rapidly spinning black holes suitable for strong-field tests of the no-hair theorem,'' \emph{Physical Review D}, vol. 83, Article ID 124015, 2011.

\bibitem{CH}
N. A. Collins and S. A. Hughes,T. Johannsen and D. Psaltis,``Towards a formalism for mapping the spacetimes of massive compact objects: Bumpy black holes and their orbits,'' \emph{Physical Review D}, vol. 69, Article ID 124022, 2004.


\bibitem{VH}
S. J. Vigeland and S. A. Hughes,``Spacetime and orbits of bumpy black holes,'' \emph{Physical Review D}, vol. 81, Article ID 024030, 2010.

\bibitem{VYS}
S. Vigeland, N. Yunes, and L. Stein,``Bumpy black holes in alternative theories of gravity,'' \emph{Physical Review D}, vol. 83, Article ID 104027, 2011.

\bibitem{KZ}
R. Konoplya and A. Zhidenko,``Detection of gravitational waves from black holes: Is there a window for alternative theories?'' \emph{Physics Letters B}, vol.750,pp.350-353, 2016.

\end{thebibliography}
\end{document}